\documentclass[conference]{IEEEtran}
\IEEEoverridecommandlockouts
\usepackage{amsmath,amssymb,amsfonts}
\usepackage{amsthm}
\usepackage{graphicx}
\usepackage{textcomp}
\usepackage{mathptmx}
\usepackage{amsmath}
\usepackage{tabularx}
\usepackage{subfigure}
\usepackage{xcolor}
\usepackage{cite}
\usepackage{multirow}
\usepackage{url}
\usepackage{float}
    
\graphicspath{{./images/}}

\usepackage{url} 
\usepackage{algorithm,algpseudocode}

\def\BibTeX{{\rm B\kern-.05em{\sc i\kern-.025em b}\kern-.08em
    T\kern-.1667em\lower.7ex\hbox{E}\kern-.125emX}}

\begin{document}

\title{\huge{VaxEquity: A Data-Driven Risk Assessment and Optimization Framework for Equitable Vaccine Distribution}
}

\author{Navpreet Kaur, Jason Hughes, and Juntao Chen
\thanks{The authors are with the Department of Computer and Information Sciences, Fordham University, New York, NY, 10023 USA. E-mail: \{nkaur15,jhughes50,jchen504\}@fordham.edu.}
\thanks{This research was supported in part by a Faculty Research Grant from Fordham Office of Research.}\vspace{-2mm}}

\maketitle

\begin{abstract}
With the continuous rise of the COVID-19 cases worldwide, it is imperative to ensure that all those vulnerable countries lacking vaccine resources can receive sufficient support to contain the risks. COVAX is such an initiative operated by the WHO to supply vaccines to the most needed countries. One critical problem faced by the COVAX is how to distribute the limited amount of vaccines to these countries in the most efficient and equitable manner. This paper aims to address this challenge by first proposing a data-driven risk assessment and prediction model and then developing a decision-making framework to support the strategic vaccine distribution. The machine learning-based risk prediction model characterizes how the risk is influenced by the underlying essential factors, e.g., the vaccination level among the population in each COVAX country. This predictive model is then leveraged to design the optimal vaccine distribution strategy that simultaneously minimizes the resulting risks while maximizing the vaccination coverage in these countries targeted by COVAX. Finally, we corroborate the proposed framework using case studies with real-world data. 
\end{abstract}

\begin{IEEEkeywords}
Pandemic Risk Assessment, Vaccine Distribution, Equity, Learning and  Optimization
\end{IEEEkeywords}

\section{Introduction}
The COVID-19 pandemic arose and spread at unexpected rates causing a great disrupt to everyday life all around the world. The pandemic highlighted many shortcomings such as shortages in medical supplies and the urgent need for the development of a vaccination. The rapid spread of the COVID-19 virus made it difficult to keep up with the demand for vaccines, especially for lower income countries. With the lack of supply for vaccine dosages and the high demand in 2021, it is difficult to ensure that vaccine allocation will be efficient to contain the spread of the virus and equitable to different countries with heterogeneous needs \cite{del2021fair}.

For the purpose of this work, we focus on vaccine distribution to COVAX countries. COVAX (COVID-19 Vaccines Global Access) is an initiative under the World Health Organization (WHO) that focuses on distributing the COVID-19 vaccines to low-to-median-income countries around the world \cite{COVAXexplained}. This initiative aims to distribute vaccines quickly and ensure that low-to-median-income countries receive a considerable amount of support during the pandemic as compared to those countries having abundant vaccine resources.

The WHO's current framework for vaccine allocation consists of distributing the vaccines in phases due to the limited supply; the initial distribution aims to cover 20\% of the population for each COVAX country and then continue to increase that percentage as the production of vaccines increases \cite{WHOequitableallocation}. With limited vaccines, it is urgent to devise an allocation plan that is both effective and fair to contain the virus in all those countries targeted by COVAX \cite{herzog2021covax,emanuel2020ethical}. In addition, the allocation strategy should consider that some countries have a higher risk of virus spreading, and some countries already have a portion of population vaccinated. Thus, the vaccine allocation mechanism needs to be adaptive to the changing situations to those targeted countries.

In this paper, we aim to develop an equitable and efficient vaccine allocation scheme for COVAX countries for which we have data available. The vaccine allocation can be seen as a decision-making problem over a network that consists of COVAX countries (vaccine recipients) and the WHO (vaccine supplier). The efficiency of the allocation scheme ensures that the vaccine distribution minimizes the health risks across all targeted countries. However, the most efficient allocation may not be the most equitable to resource recipients, a phenomenon commonly observed in resource allocation problems \cite{hughes2021fair}. When merely considering efficiency, it is possible that only several countries receive most of the vaccine resources. Thus, it is imperative to consider the equity in the vaccine distribution scheme \cite{national2020framework}, in which one needs to incorporate the vaccination rate in each country into the decision-making framework. 

To achieve this goal, our first step is to gather and analyze COVID-19 data from the COVAX countries. The data-driven risk assessment yields formal risk metrics. We then resort to machine learning techniques to learn the relationship between the risk outcome and their features including vaccination rate, death rate, etc. The learned models are powerful to predict the risks under different circumstances which is essential for the vaccine distribution. Through the predictive model, we then establish an optimization problem in which the objective function includes the aggregated risks and the vaccination levels of all targeted countries. Specifically, the first part of the objective captures the efficiency while the second part promotes the equity of vaccine distribution. The proposed decision-making framework shows promising results using the collected data for the COVAX countries.

The rest of the paper is organized as follows. Section \ref{sec:risk_assessment} presents the essential steps for data pre-processing and defines the risk assessment metric. Section \ref{sec:learningrisk} develops and evaluates a number of machine learning models for risk quantification and prediction. From there, we develop a mathematical framework to enable fair and effective vaccine allocation in Section \ref{sec:framework}. Section \ref{sec:results} corroborates the proposed schemes using case studies. Finally, Section \ref{sec:conclusion} concludes the paper.

\section{Data-Driven Risk Assessment}\label{sec:risk_assessment}
This section first describes the essential steps in the data pre-processing and then develops a metric to quantify the evolving health risk of those targeted countries.

\subsection{Data Pre-Processing}
The data collection consists of records from 31 COVAX countries, and the dataset is available at \textit{Our World in Data} \cite{owidcoronavirus}. The values for each country include time series records for daily new cases, total number of cases, daily new deaths, total number of deaths, people vaccinated, people fully vaccinated, and total vaccinations per day. The numbers of people vaccinated, people fully vaccinated, and total vaccinations are accumulated over days. In the modeling, we focus on the people vaccinated attribute. Our data also includes population, hospital beds available per one thousand people, and the human development index for each country. The human development index is a measurement of human development using factors such as the standard of living and education \cite{owidhumandevelopmentindex}. Human development index and hospital beds available are factors that do not immensely influence our model, however, they provide us with an understanding of why certain countries with similar economic standards may have different responses to the pandemic. The data is cleaned by removing values that are illogical, such as negative case counts and filling empty data points with values of zero. The data values report records from the point where COVID-19 was first detected (around February 2020) to the end of November 2021.

\subsection{Risk Assessment}
Using the collected data, we can quantify the death rate, vaccination rate, and the risk metric. We follow the framework published by the Center of Disease Control and Prevention's (CDC) in which the risk metric is defined to be based on the new cases values from the dataset \cite{WHOequitableallocation}. Specifically, we take the sliding window average over 28 days of new cases and divide it by the population to obtain the risk and then normalize it. In other words, the risk on day $t$, $R(t)$, admits a following quantification:
\begin{equation}
    R(t) = \frac{\sum_{k=t-27}^t NewCase(k)}{28\cdot Population},
\end{equation}
where $NewCase(k)$ denotes new cases on day $k$. The values for death rate and vaccination rate are obtained by taking their values for each day and dividing by the population. 

\begin{figure}[!t]
\centering
    \includegraphics[width=.8\columnwidth]{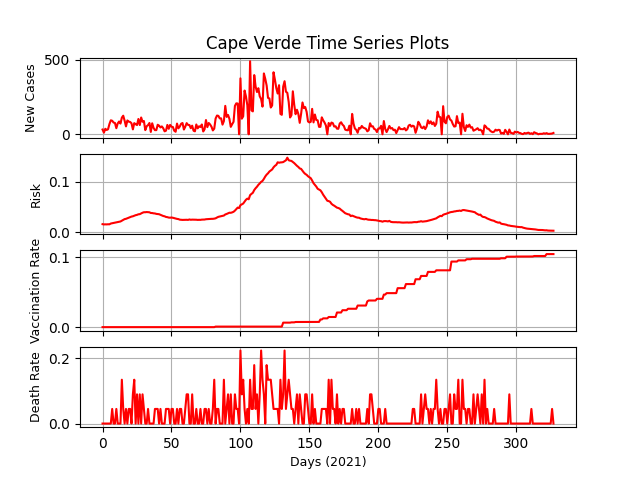}
    \caption{New cases, vaccination rate, death rate, and the health risk for Cape Verde in 2021.}
    \label{fig:data}
    \vspace{-4mm}
\end{figure}

\begin{figure}[!t]
    \centering
    \subfigure[Ghana]{
    \includegraphics[width=.45\columnwidth]{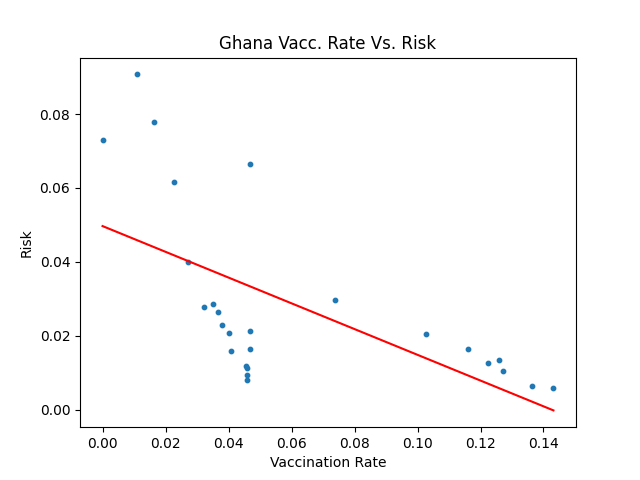}}
    \subfigure[ Madagascar]{
    \includegraphics[width=0.45\columnwidth]{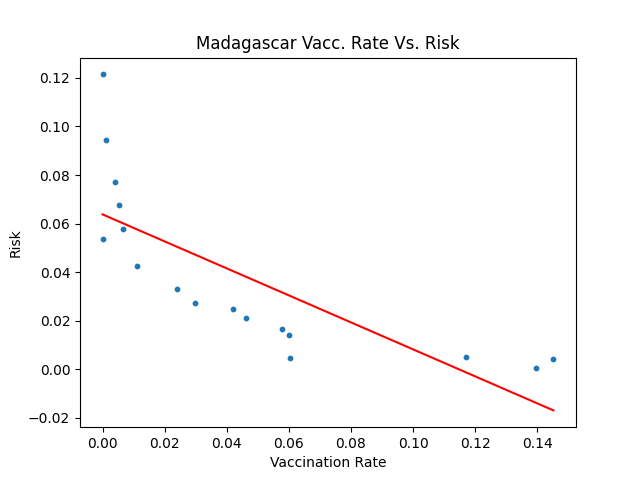}}
    \caption{Relationship between vaccination rate and risk. It can be observed that the risk decreases as the vaccination rate increases among the population.}
    \label{fig:vaccine_risk}
    \vspace{-4mm}
\end{figure}

In Fig. \ref{fig:data}, we illustrate the raw data values (e.g., daily new cases) and calculated values (e.g., vaccination rate) for Cape Verde. The time series plots depict the general trends and relations between our chosen features. The `New Cases' subplot shows the increases in the first days of the 2021 year, but after about 130 days into the year they begin to decrease slowly. We deduce that this is due to the emergence of vaccines and people taking the vaccines in Cape Verde, since new case values begin to decrease at around the same time where the subplot for the vaccination rate begins to increase. To examine the relationship between the vaccination rate and the risk, we depict the corresponding results for Ghana and Madagascar in Fig. \ref{fig:vaccine_risk}. 
The results indicate that there is a negative correlation between the vaccination rate of a country and its pandemic risk. 
Specifically, as the vaccination rate begins to increase, risk begins to decrease significantly with a corresponding rate. The inherent correlation can be approximately linear over the interested parameter regime as shown in Fig. \ref{fig:vaccine_risk}. 

As mentioned earlier, the data source contains information from the first detection of the COVID-19 virus in the country to November 2021. Our models are tested for two different scenarios: one in which the data points for years 2020 and 2021 are split, and the other in which they are not. Due to the fact that none of the COVAX countries received the vaccine prior to 2021, generating models on the year 2021 alone is more logical. We corroborate this idea by testing both cases, and conclude that the results yielded by the data points from only year 2021 are more accountable and generalizable.

\section{Learning the Risk Model}\label{sec:learningrisk}
In this section, we aim to construct a model that predicts the evolving risks using machine learning approaches. The learned model will be leveraged to determine the vaccine allocation schemes in the next section. To achieve the goal, three different models are investigated: artificial neural network (ANN), convolutional neural network (CNN), and linear regression (LR) \cite{bishop2006pattern}. For each model, we denote the set of training features by $X$ and the targeted variable by $Y$, where $X$ = [normalized death rate, normalized vaccination rate, hospital beds available per thousand, human development index] and $Y$ = [normalized risk metric]. We train these three models with $X$ as the input features to most accurately predict the target $Y$.

In terms of the selected features, the death rate can be an indicator of how contagious the virus is, which influences the new infections. The vaccination rate plays an important role in combating the virus spreading and hence mitigates the risk. The vaccination and death rates may have inherent correlation but its extent is unclear based on the currently available statistics (knowing that the death due to COVID-19 depends on many other factors, such as treatment and medical history). Thus, we incorporate both death and vaccination rates as features. We admit that the risk also depends on other social, political, and informational factors, and they will be investigated in detail in subsequent works.

\subsection{Artificial Neural Network}
We use an ANN model consisting of 4 hidden layers together with the input and output layers, and the model has a total of 512 nodes. We split the dataset with 95\% used for training and the remaining 5\% used for testing purposes. The learning accuracy is evaluated by the  mean absolute error (MAE), mean squared error (MSE), mean percentage error, and mean percentage difference. The results are summarized in Table I. It can be seen that the performance of the trained model is satisfying as the values calculated are decently small. To account for any noise and overfitting, we further test a model with a dropout function included. However, there is no significant improvement on the performance so it is omitted.

\subsection{Convolutional Neural Network}
We further investigate a CNN model with a total of 5 layers and 256 nodes. The test size and training size for the CNN model are the same as the ANN model. The accuracy of the CNN model is also evaluated with regard to the same metrics and the results are shown in Table I. To account for any noise and overfitting, a dropout function is added in the model during the training. The results indicate that the CNN model also achieves a good performance.

\subsection{\textit{Linear Regression}}
This section studies the LR model for risk prediction. 
Specifically, the constructed LR model admits the following form:
\begin{equation}\label{LR_eqn}
\begin{aligned}
    R_{j}(t) = \beta_{0j} + \beta_{1j} D_{j}(t) + \beta_{2j} V_{j}(t)
    +\beta_{3j} H_{j}(t) + \beta_{4j} I_{j}(t),
\end{aligned}
\end{equation}
where $ R_{j}(t) $ is the predicted risk metric;
      $ V_{j}(t) $ is the vaccination rate;
      $ D_{j}(t) $ is the death rate;
      $ H_{j}(t) $ is the hospital beds available per thousand (which is time-invariant);
      $ I_{j}(t) $ is the human development index (which is time-invariant); and all these variables are
      associated with the $j$th country in the set of COVAX countries on day $t$. In addition, $\beta_{0j}, \beta_{1j}, \beta_{2j}, \beta_{3j},$ and $\beta_{4j}$ are coefficients associated with the corresponding feature variables for the targeted country $j$.

The training and testing sets are split with a 70\%/30\% scale. Similarly, MAE, MSE, and r-squared values are computed for performance evaluation. Table II shows the obtained results. It can be concluded that the LR model is efficient in predicting the risks. In comparison to the ANN and CNN model, LR has a slightly larger error, but it also yields sufficiently good performance and is adequate to be used in the decision-making framework for vaccine allocation in Section \ref{sec:framework}.

\begin{table}[!t]
\begin{center}
\begin{tabular}{||c | c | c ||}
    \hline
    Metric & ANN & CNN  \\ [0.5ex]
    \hline\hline
    MAE Average & 0.0200  & 0.0150 \\
    \hline
    MAE Median & 0.0184 & 0.0131 \\
    \hline
    MSE Average & 0.0012 & 0.0009 \\
    \hline
    MSE Median & 0.0007 & 0.0004 \\ 
    \hline
    Mean \% Diff. Average & 0.7702 & 0.5455 \\
    \hline
    Mean \% Diff. Median & 0.6829 & 0.4124 \\
    \hline
\end{tabular}
\end{center}
\caption{Performance results of the neural network models. }
\end{table}

\begin{table}[!t]
\begin{center}
\begin{tabular}{|| c | c ||}
    \hline
    Metric & LR \\ [0.5ex]
    \hline\hline
    MAE Average & 0.0214\\
    \hline 
    MAE Median & 0.0224\\
    \hline
    MSE Average & 0.0009\\
    \hline
    MSE Median & 0.0008\\
    \hline
    $R^2$ Average & 0.4189\\
    \hline
    $R^2$ Median & 0.4189\\
    \hline
\end{tabular}
\end{center}
\caption{Performance results of the linear regression model.}
\vspace{-5mm}
\end{table}

\begin{figure}[!h]
  \centering
  \subfigure[Results of LR model for Indonesia]{
    \includegraphics[width=0.45\columnwidth]{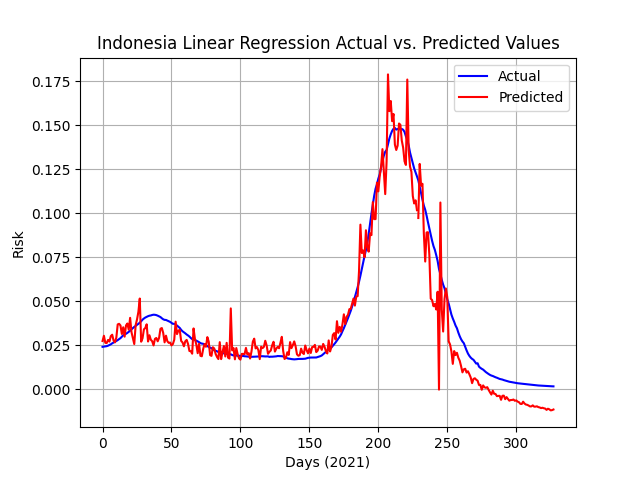}\label{fig:Reg1}}
	 \subfigure[Results of LR model for Malawi]{
    \includegraphics[width=0.45\columnwidth]{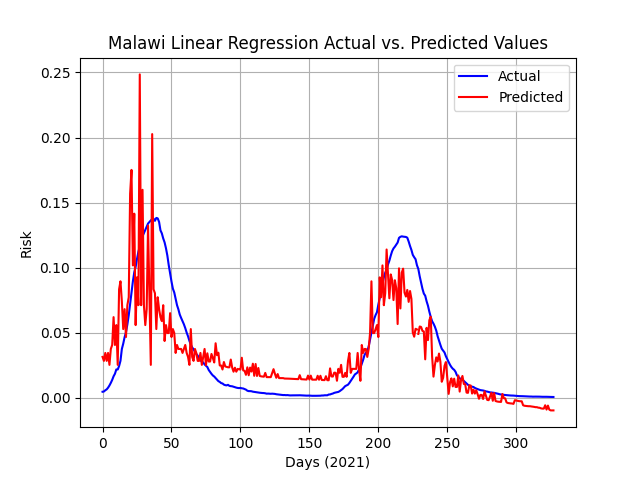}\label{fig:Reg2}}
  \caption[]{Risk prediction performance using LR model for Indonesia and Malawi. The prediction errors in both cases are relatively low.}
  \label{fig:regressionModels}
\end{figure}

Despite the ANN and CNN model having relatively better results, the LR model is more explainable with a satisfying accuracy for our implementation stage. The linear model gives us sufficient accuracy over the interested parameter regime. Fig. \ref{fig:regressionModels} shows the performance of our model by presenting the actual risk and the predicted risk for two countries, Indonesia and Malawi. It can be seen that the model yields satisfying results in which the prediction error is relatively small. Thus, our risk prediction model is sufficiently accurate for predicting the risk for the list of targeted COVAX countries.

\section{Mathematical Framework for Equitable Vaccine Allocation}\label{sec:framework}
The previous sections have developed a data-driven risk quantification and prediction model, and characterized how the vaccine contributes to mitigating the health risks in COVAX countries.
This section will leverage the obtained model to develop a decision-making framework for equitable and efficient distribution of limited vaccines to those targeted countries.

Based on the LR model in \eqref{LR_eqn}, we construct the following metric to capture the aggregated risk of all targeted countries:
\begin{equation}\label{risk_sum}
\sum_{j \in \mathcal{J}} R_{j}(t) = \sum_{j \in \mathcal{J}} (\tilde{\beta}_{0j} + \beta_{1j}  D_{j}(t) + \beta_{2j}  V_{j}(t)),
\end{equation}
where $\mathcal{J}:=\{1,2,...,J\}$ is the list of COVAX countries. Note that the only change in this function is that the terms associated with the coefficients for hospital beds available per thousand and human development index are combined with $ \beta_{0j} $ since those two components are constant based on the learnt model. Equivalently, we have $\tilde{\beta}_{0j} = \beta_{0j}+\beta_{3j} H_{j}(t) + \beta_{4j} I_{j}(t)$ in \eqref{risk_sum}. Based on the learning results, we have the following observation on the coefficients (shown in Fig. \ref{Coefficients} later): $\beta_{0}$ is relatively small, whereas $\beta_{1}$ and $\beta_{2}$ indicate a strong positive and negative relation, respectively, with the risk outcome.

Our goal is to minimize the risk of COVID-19 across the targeted countries, captured by \eqref{risk_sum}, by developing an optimal vaccine allocation scheme. In order to minimize such a risk, we need to control $ V_{j}(t) $, and all other variables are provided to us from the data. In this work, we consider a scenario that the vaccines are distributed on day $t$ to these countries based on their real-time situation in terms of vaccination rate, death rate, etc. Here, we do not specify a fixed $t$, and thus our framework is flexible with a plug-and-play feature for any day. To minimize the aggregated risk, we need to solve the following optimization problem:
\begin{equation} \label{eqn:OP}
\begin{aligned} 
\mathrm{(OP):}\ \min_{V_{j}(t),\forall j\in \mathcal{J}}&\quad \sum_{j \in \mathcal{J}} R_{j}(t)\\
\mathrm{s.t.}\quad  &0 \leq \sum_{j \in \mathcal{J}} (V_{j}(t) - V_{j}(t-1))  P_j \leq TV,\\
&0\leq V_{j}(t)\leq 1,\ \forall j\in\mathcal{J},
\end{aligned}
\end{equation}
where $P_j$ is the population of the $j$th country, and $TV$ is the total amount of vaccines available for distribution for the COVAX countries. The first constraint indicates that the total distributed vaccines is upper bounded by $TV$. The second constraint is a natural bound for the vaccination rate. After obtaining $V_j(t),\ \forall j\in\mathcal{J}$, we then obtain that $(V_{j}(t) - V_{j}(t-1))  P_j$ vaccine doses should be allocated to country $j\in\mathcal{J}$.

One drawback of the optimization framework \eqref{eqn:OP} is that it does not give an equitable solution. It can be observed that the resulting vaccine allocation scheme will allocate most of the resources to a small set of countries where the vaccines are comparatively effective to mitigate the risk (i.e., those countries with a smaller $\beta_{2j}$).  To achieve an equitable vaccine allocation, we incorporate an additional fairness metric to the objective function other than the risk. The optimization problem for fair vaccine allocation is presented as follows:
\begin{equation} \label{eqn:equation_name} 
\begin{aligned} 
\mathrm{(OP-Fair):}&\\
\min_{V_{j}(t),\forall j\in \mathcal{J}}&\quad \sum_{j \in \mathcal{J}} R_{j}(t) - \omega \frac{(\sum_{j \in \mathcal{J}} V_{j}(t))^2}{J\sum_{j \in \mathcal{J}} V_{j}(t)^2}\\
\mathrm{s.t.}\quad  &0 \leq \sum_{j \in \mathcal{J}} (V_{j}(t) - V_{j}(t-1))  P_j \leq TV,\\
&0\leq V_{j}(t)\leq 1,\ \forall j\in\mathcal{J},
\end{aligned}
\end{equation}
where $\omega \geq 0$ is a weighing constant for fairness and $J=31$ as there are in total 31 countries considered in our framework. The fairness term $\frac{(\sum_{j \in \mathcal{J}} V_{j}(t))^2}{J\sum_{j \in J} V_{j}(t)^2}$ is based on the Jain's fairness index \cite{jain1984quantitative,lan2010axiomatic}. One advantage of this fairness term is that the resulting solution takes into account the vaccination level of each individual country. Intuitively, the planner will distribute the vaccine resources with priority to those countries with a low level of vaccination rate (i.e., generally the countries with a significant shortage of vaccines), while considering its trade-off against a high-level of combined risk outcome across all targeted COVAX countries. Therefore, we can choose an appropriate $\omega$ to balance efficiency and fairness in the vaccine allocation.

\section{Case Studies and Discussions}\label{sec:results}
In this section, we corroborate the developed framework for equitable vaccine allocation using case studies. 

\subsection{Setup and Discussion on the Learned Risk Model}
The WHO planned to allocate $300,000,000$ vaccine doses to COVAX countries. We assume that these vaccines will be distributed over a 100 day period (may not be consecutive), and thus a total of $TV=3,000,000$ vaccines is scheduled for allocation per each distribution day. In addition, we select $t = 100$ which is the day on which vaccines are distributed. Note that the values of $TV$ and $t$ can be chosen according to real situations. Though our current result does not include every country enlisted as COVAX, proper data from the remaining COVAX countries can be added to the allocation framework in the implementation.

\begin{figure}[!t]
  \centering
  \subfigure[Population of the targeted COVAX countries.]{
    \includegraphics[width=0.45\columnwidth]{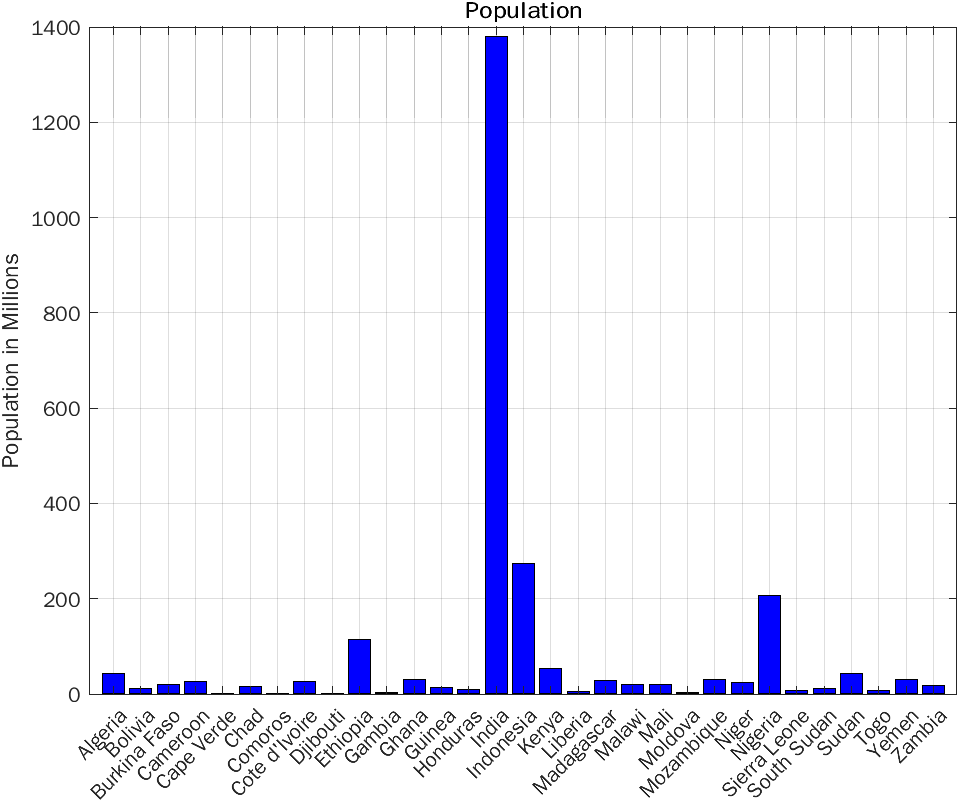}\label{Population}}
	 \subfigure[Coefficients in the learnt risk function.]{
    \includegraphics[width=0.45\columnwidth]{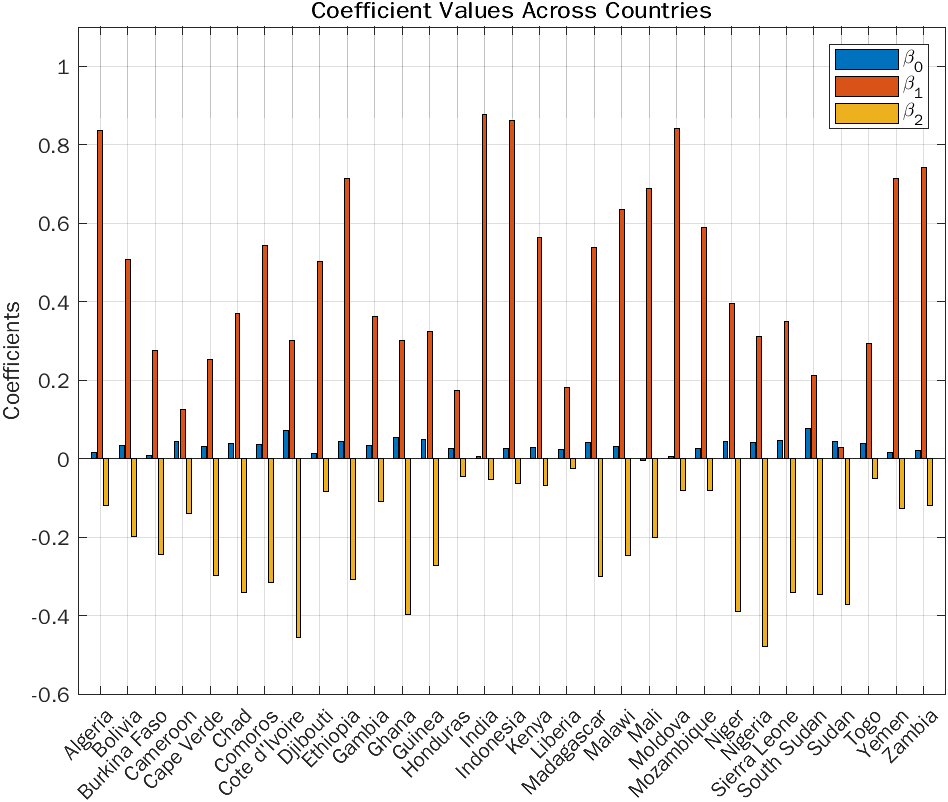}\label{Coefficients}}
  \caption[]{Population and risk function coefficients utilized in the model.}
  \label{fig:Parameters}
\end{figure}

Fig. \ref{Population} depicts the population statistics for the targeted countries to which the vaccines will be distributed. Note that population is a parameter which is utilized in evaluating various metrics such as the vaccination rate, death rate, and health risk. 
It hence plays an important role in determining an equitable strategy for vaccine distribution among COVAX countries. It is reasonable to conjecture that countries with higher populations, e.g., India and Indonesia, would require a larger supply of vaccines, considering that all the countries in our framework are low-income countries with similar conditions of fighting against the pandemic. We can predict the health risks using the constructed LR model \eqref{LR_eqn}. The coefficients $\beta_{0}, \beta_{1},$ and $\beta_{2}$ in the learned model for each country are shown in Fig. \ref{Coefficients}. 
The values of $\beta_{0}$ are extremely close to 0, showing that they do not significantly impact the risk quantification. 
The values of $\beta_{1}$, on the other hand, are much more prominent, with most coefficient values above 0.5 for each country. $\beta_{1}$ is the coefficient associated with the death rate and the result indicates a positive relation between this factor and the risk output. This is explainable as an increasing death toll indicates the virus being more contagious, resulting in more infected people and hence a larger risk.  
In comparison, $\beta_{2}$ has a negative correlation with the risk. The coefficient values of $\beta_{2}$ associated with the vaccination rate are all negative. This relation concurs with the fact that due to the introduction of the vaccination and its success in countering the COVID-19 virus, the new cases reported each day began to decrease as more and more people got vaccinated. 

\begin{figure}[!t]
  \centering
  \subfigure[Vaccination rate after the proposed scheme with and without fairness in the implementation.]{
    \includegraphics[width=0.45\columnwidth]{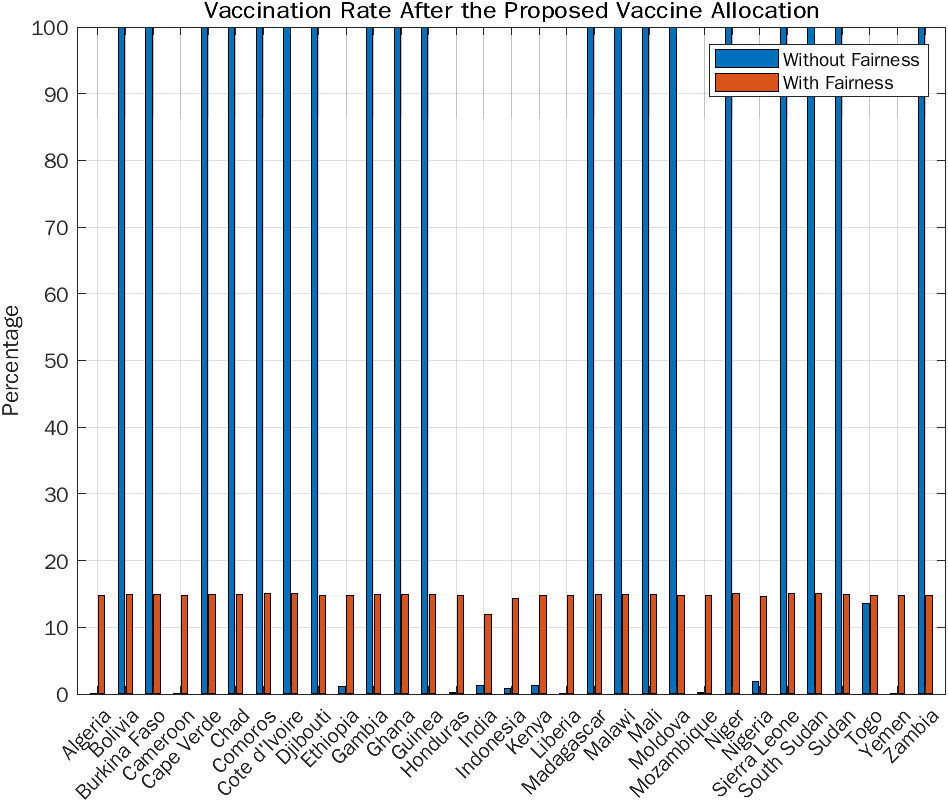}\label{fig:VaccRateFair}}
	 \subfigure[Vaccination allocation scheme with and without fairness in the implementation.]{
    \includegraphics[width=0.45\columnwidth]{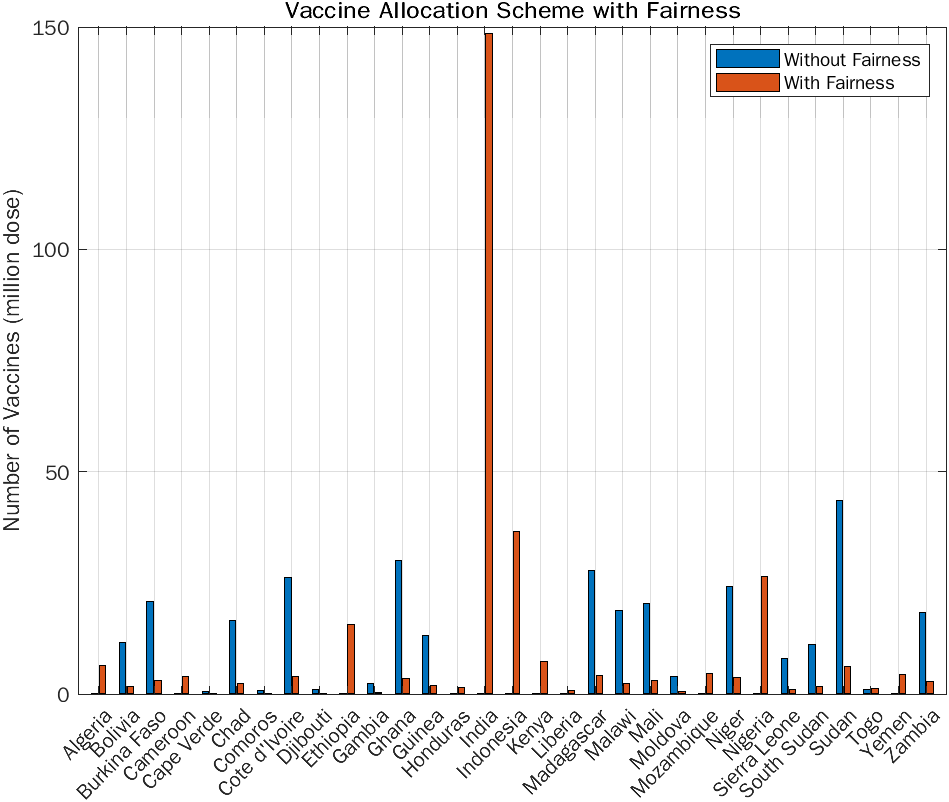}\label{fig:VaccineAllocFair}}
  \caption[]{Results of the proposed scheme with and without fairness incorporated into decision-making of vaccine distribution.}
  \label{fig:WithandWithoutFairness}
\end{figure}


It is expected that as the vaccination rates for countries increase, there should be a reduction in the pandemic risk.
We next investigate the vaccine distribution scheme yielded by the proposed optimization framework.

\subsection{Equitable vs. Inequitable Vaccine Distribution}
We first study how the vaccine allocation plan changes due to the inclusion of fairness into decision-making. Fig. \ref{fig:WithandWithoutFairness} illustrates the obtained results by solving (OP-Fair). Specifically, Fig. \ref{fig:VaccRateFair} shows the vaccination rates for all considered countries after the proposed vaccine distribution schemes shown in Fig. \ref{fig:VaccineAllocFair} under two scenarios with $\omega=0$ (no fairness) and $\omega=50$ (with fairness). When fairness is not considered, i.e., $\omega=0$, the results demonstrate that countries with low populations are allocated enough vaccines for achieving 100\% vaccination rates. In comparison, countries with much larger populations, such as India and Indonesia, are not allocated with sufficient amount of vaccines and will end up with extremely low vaccination rates. Such a distribution scheme is not equitable considering that not all countries are treated the same and do not get enough of their country vaccinated whereas other countries achieve much higher vaccination rates. 
To remedy this consequence, the scenario with $\omega=50$ promotes fairness in vaccine distribution explicitly. Though the vaccination rates in this case become smaller as shown in Fig. \ref{fig:VaccRateFair}, this distribution strategy is better suited for a fair distribution. Under the proposed plan, most countries in the framework attain a similar vaccination rate. The result shown in Fig. \ref{fig:VaccineAllocFair} indicates that the equitable distribution plan specifically considers the population of each country, other than the effectiveness of vaccines in mitigating the risks.

\begin{figure}[!t]
  \centering
  \subfigure[Vaccination rate until 4/10/2021.]{
    \includegraphics[width=0.45\columnwidth]{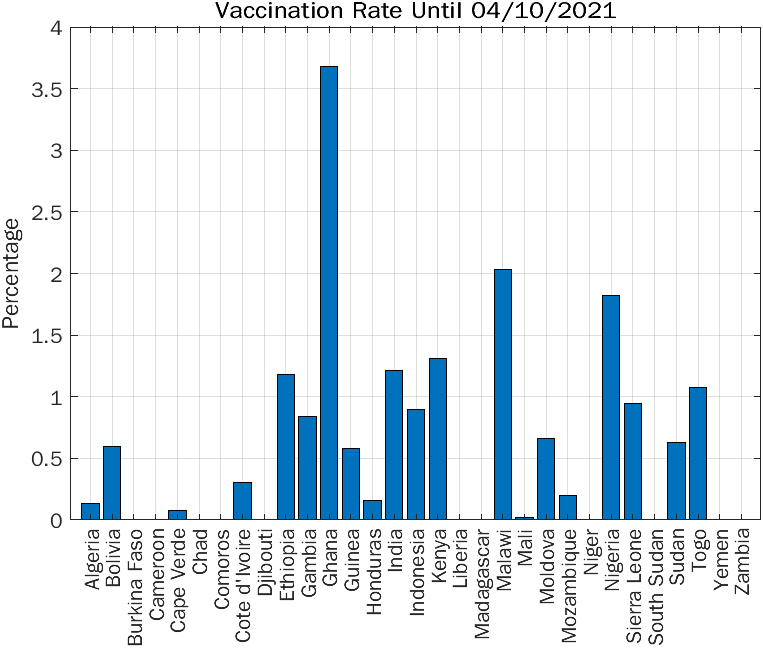}\label{fig:Vaccination_rate_until_before_410}}
	 \subfigure[Vaccination rate after the proposed scheme.]{
    \includegraphics[width=0.45\columnwidth]{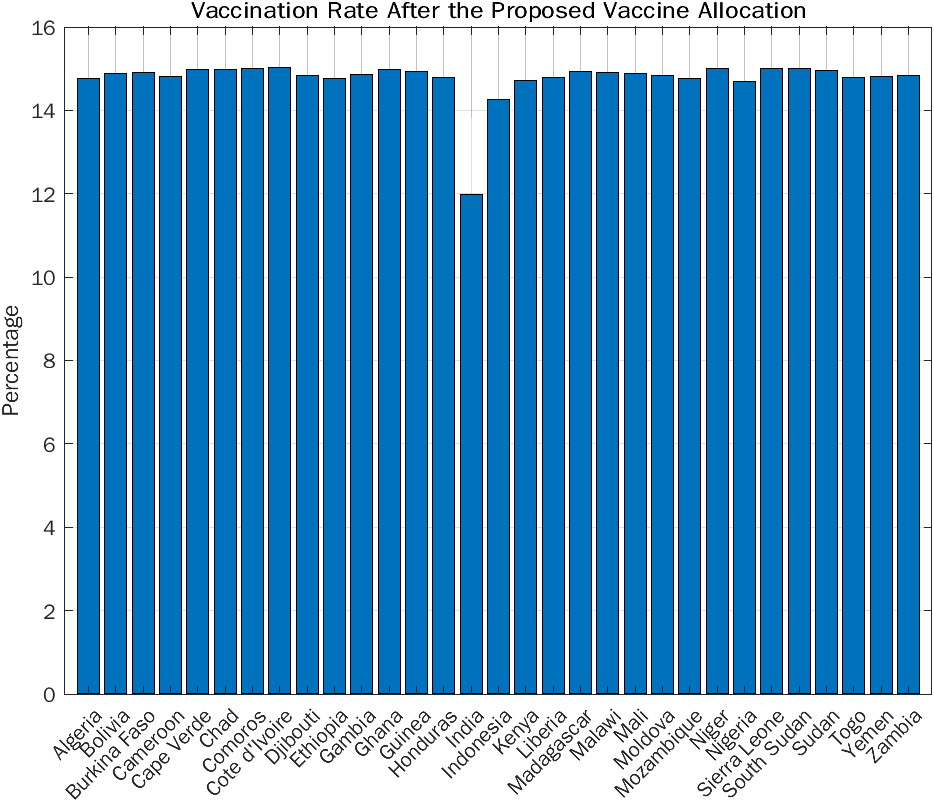}\label{fig:Vaccination_rate_after_the_proposed_vaccine_allocation_scheme}}
    \subfigure[Vaccination allocation after the proposed scheme.]{
    \includegraphics[width=0.45\columnwidth]{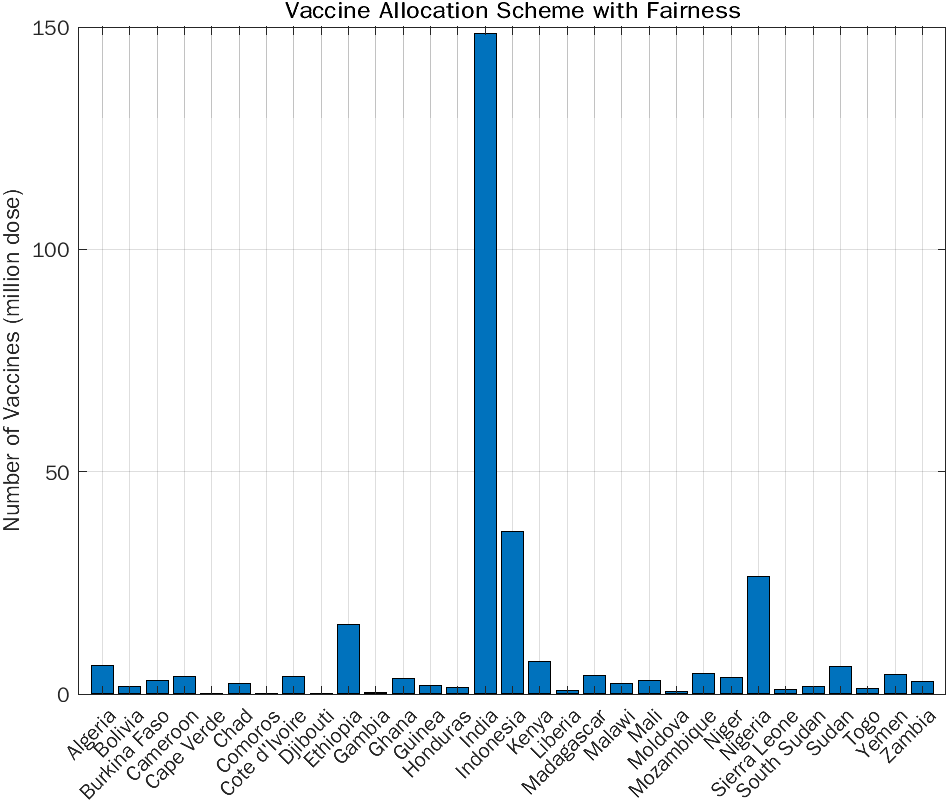}\label{fig:Vaccine_allocation_after_the_proposed_scheme}}
    \subfigure[Risk reduction after the proposed scheme.]{
    \includegraphics[width=0.45\columnwidth]{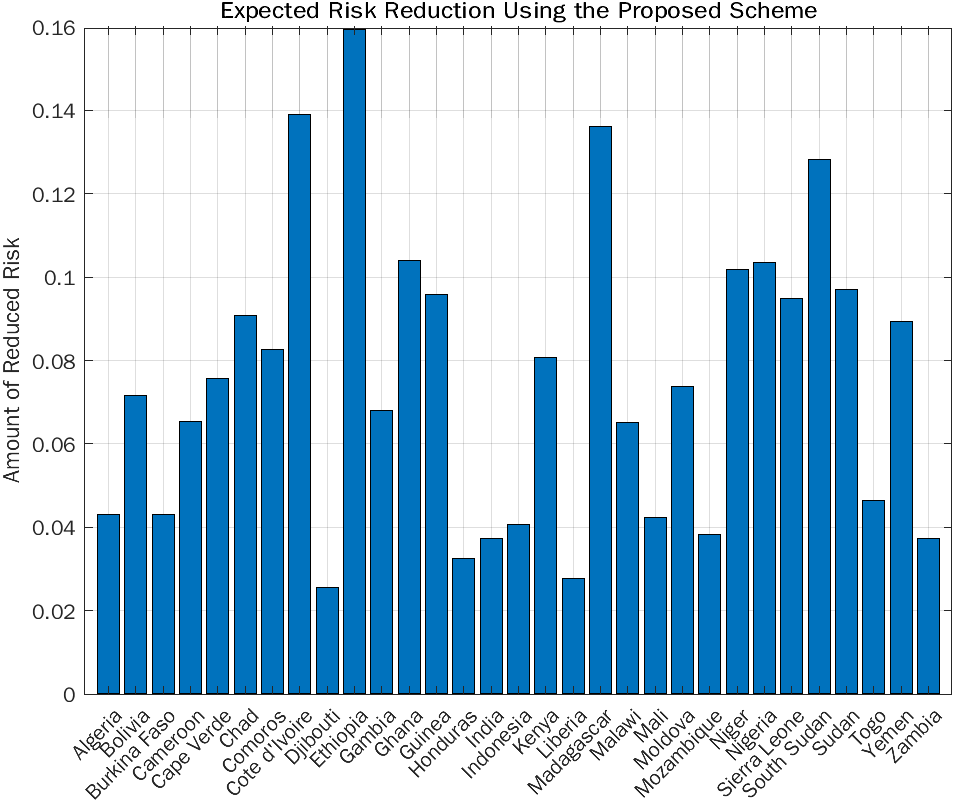}\label{fig:Expected_risk_reduction}}
  \caption[]{Equitable vaccine allocation scheme with $\omega=50$.}
  \label{fig:VaccSchemeWithOmega50}
\end{figure}

To demonstrate the equitable framework further, we refer to Fig. \ref{fig:VaccSchemeWithOmega50}, which are the results corresponding to $\omega = 50$. Due to the assumption that the vaccine distribution of $TV=300,000,000$ vaccines happens over a span of 100 days, we assume 4/10/2021 as a distribution day and plot what the actual vaccination rate has been up till this date in Fig. \ref{fig:Vaccination_rate_until_before_410}. The bars indicate different vaccine rates all across the countries, which is a fact needed to be taken into account when developing a new distribution scheme. Fig. \ref{fig:Vaccination_rate_after_the_proposed_vaccine_allocation_scheme} depicts the attainable vaccination rates under the proposed equitable scheme shown in Fig.  \ref{fig:Vaccine_allocation_after_the_proposed_scheme}. This distribution is considered equitable due to the inclusion of the varying risks and population sizes in the proposed framework in (OP-Fair). 
All countries in the considered network receive an equitable amount of vaccines. To evaluate the efficacy of the proposed model, we calculate the amount of potential risk reduction for each country under the proposed vaccination distribution and the result is plotted in Fig. \ref{fig:Expected_risk_reduction}. We can observe that all countries exhibit a certain level of risk reduction. Though countries like Djibouti and India reveal lower levels of risk reduction compared to the other countries, they possess different factors that may influence the effectiveness of the vaccines in their countries. Such factors include whether these countries are implementing strict social distancing rules and the government's policies towards vaccine distribution to their populations at various scales (e.g., state-level, city-level, community-level, etc).

\begin{figure}[!t]
  \centering
  \subfigure[Vaccination rate after the proposed scheme.]{
    \includegraphics[width=0.45\columnwidth]{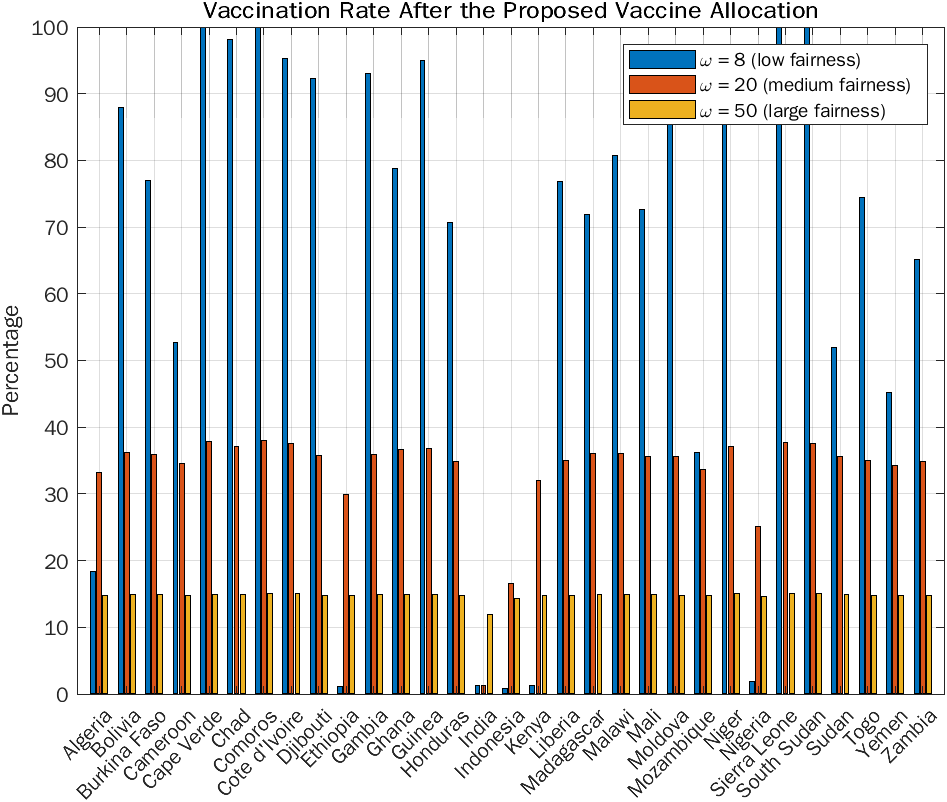}\label{fig:omega_vacc_rate}}
	 \subfigure[Vaccination allocation scheme.]{
    \includegraphics[width=0.45\columnwidth]{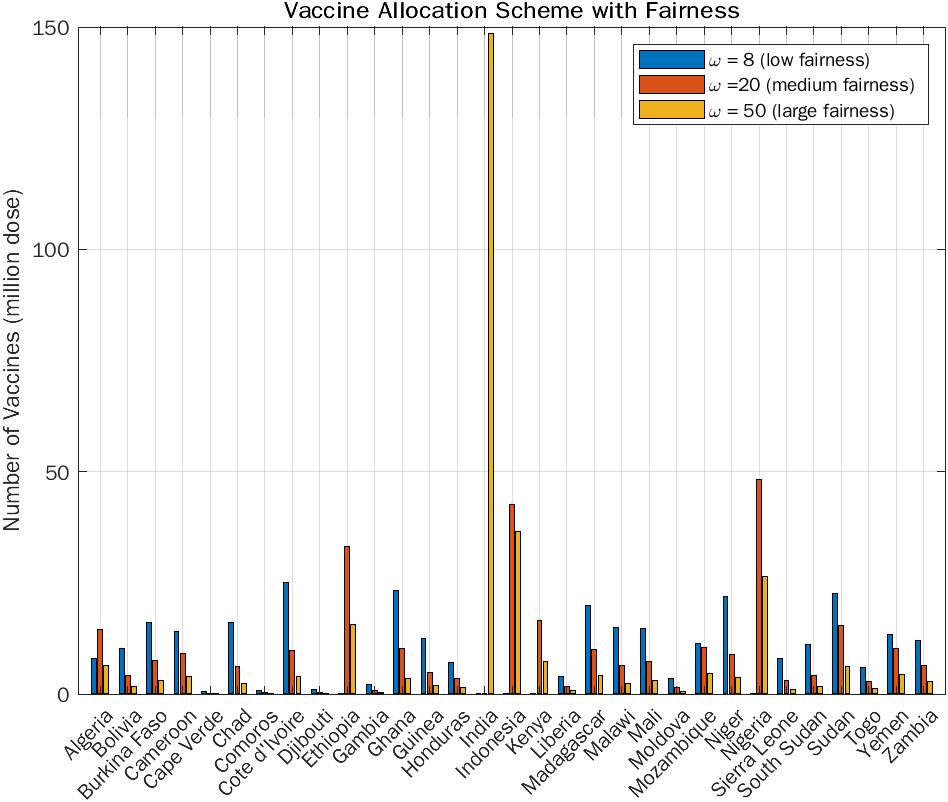}\label{fig:omega_vacc_alloc}}
    \subfigure[Risk reduction.]{
    \includegraphics[width=0.45\columnwidth]{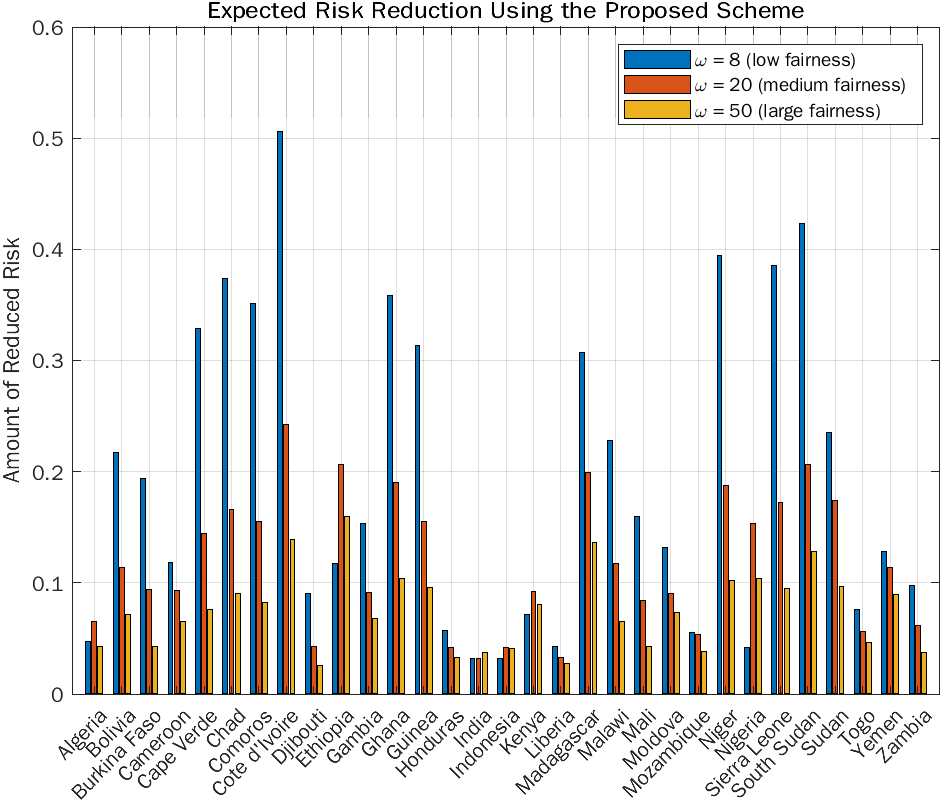}\label{fig:omega_risk_reduc}}
  \caption[]{Vaccine allocation scheme with varying level of fairness considerations ($\omega=8, \omega=20,$ and $\omega=50$). A larger $\omega$ yields a more equitable outcome.}
  \label{fig:VaccRateSchemeFair}
\end{figure}

\subsection{Impact of the Degree of Equity}
To understand the influence of equity in the resulting distribution strategy, we implement various weighing factors of $\omega$ in the developed optimization framework. This set of case studies helps to determine the degree of equity incorporated to decision-making to have a balance between efficiency and equity. Fig. \ref{fig:VaccRateSchemeFair} demonstrates the results of vaccination rate, vaccine allocation schemes, and risk reduction for various fairness levels with $\omega=8,\ \omega=20,$  and $\omega=50$. In Fig. \ref{fig:omega_vacc_rate}, when $\omega=8$, the distribution scheme is not sufficiently equitable but is the most efficient one among the three cases as it reduces the largest amount of aggregated risks of all considered countries. As the weighing factor increases, the vaccination rates get closer among different targeted countries. It can be seen that $\omega=50$ leads to the most equitable outcome, indicating that a higher weighing factor produces a more equitable allocation. Fig. \ref{fig:omega_vacc_alloc} plots the corresponding vaccine allocation schemes using these different values of $\omega$. With a larger $\omega$, the vaccine allocation plan becomes more population-aware, i.e., the strategy will pay more attention to the vaccination rate in the decision-making process. Fig. \ref{fig:omega_risk_reduc} displays the risk reduction in each proposed plan. Despite the larger risk reduction values depicted by the blue bars for certain countries, this value of $\omega$ is not ideal for equity since not all countries receive an appropriate amount of vaccines to reduce their risks. In comparison, under the more equitable plan with $\omega=50$, the variance of risk reduction across all countries becomes much smaller. A larger $\omega$ ensures the distribution plan to treat the countries fairly so that they are able to efficiently contain the pandemic risk among the population.

\section{Conclusion}\label{sec:conclusion}
In this paper, we have developed a theoretical framework for equitable and efficient COVID-19 vaccine distribution among countries targeted by COVAX. The established model explicitly considers the vaccination rates among the population and continuously evaluates the risks of virus propagation in these countries. The developed machine learning enabled risk prediction paradigm quantifies the effectiveness of increased vaccination level in mitigating the risks in each country, hence facilitating the optimal distribution of limited vaccines.
The incorporation of a fairness metric into the objective function successfully yields a distribution scheme that is equitable and effective in containing the global health risks. As for future work,
we plan to incorporate social and political factors, such as re-opening plans of cities, social distancing and mask wearing policies, and information broadcast by social media \cite{liu2021herd}, into the framework to investigate their influence on the virus propagation and develop more effective equitable vaccine distribution schemes. 

\bibliographystyle{IEEEtran}
\bibliography{IEEEabrv,references}

\begin{thebibliography}{10}
\providecommand{\url}[1]{#1}
\csname url@samestyle\endcsname
\providecommand{\newblock}{\relax}
\providecommand{\bibinfo}[2]{#2}
\providecommand{\BIBentrySTDinterwordspacing}{\spaceskip=0pt\relax}
\providecommand{\BIBentryALTinterwordstretchfactor}{4}
\providecommand{\BIBentryALTinterwordspacing}{\spaceskip=\fontdimen2\font plus
\BIBentryALTinterwordstretchfactor\fontdimen3\font minus
  \fontdimen4\font\relax}
\providecommand{\BIBforeignlanguage}[2]{{%
\expandafter\ifx\csname l@#1\endcsname\relax
\typeout{** WARNING: IEEEtran.bst: No hyphenation pattern has been}%
\typeout{** loaded for the language `#1'. Using the pattern for}%
\typeout{** the default language instead.}%
\else
\language=\csname l@#1\endcsname
\fi
#2}}
\providecommand{\BIBdecl}{\relax}
\BIBdecl

\bibitem{del2021fair}
A.~del Carmen Mungu{\'\i}a-L{\'o}pez and J.~M. Ponce-Ortega, ``Fair allocation
  of potential covid-19 vaccines using an optimization-based strategy,''
  \emph{Process Integration and Optimization for Sustainability}, vol.~5,
  no.~1, pp. 3--12, 2021.

\bibitem{COVAXexplained}
S.~Berkley, ``{COVAX} explained,'' \emph{Gavi The Vaccine Alliance}, 2020,
  \url{https://www.gavi.org/vaccineswork/covax-explained}.

\bibitem{WHOequitableallocation}
``{WHO} concept for fair access and equitable allocation of {COVID}-19 health
  products,'' \emph{World Health Organization}, 2020,
  \url{https://www.who.int/docs/default-source/coronaviruse/who-covid19-vaccine-allocation-final-working-version-9sept.pdf}.

\bibitem{herzog2021covax}
L.~M. Herzog, O.~F. Norheim, E.~J. Emanuel, and M.~S. McCoy, ``{COVAX} must go
  beyond proportional allocation of covid vaccines to ensure fair and equitable
  access,'' \emph{BMJ}, vol. 372, 2021.

\bibitem{emanuel2020ethical}
E.~J. Emanuel, G.~Persad, A.~Kern, A.~Buchanan, C.~Fabre, D.~Halliday,
  J.~Heath, L.~Herzog, R.~Leland, E.~T. Lemango \emph{et~al.}, ``An ethical
  framework for global vaccine allocation,'' \emph{Science}, vol. 369, no.
  6509, pp. 1309--1312, 2020.

\bibitem{hughes2021fair}
J.~Hughes and J.~Chen, ``Fair and distributed dynamic optimal transport for
  resource allocation over networks,'' in \emph{55th Annual Conference on
  Information Sciences and Systems (CISS)}, 2021, pp. 1--6.

\bibitem{national2020framework}
{National Academies of Sciences, Engineering, and Medicine}, \emph{Framework
  for Equitable Allocation of COVID-19 Vaccine}.\hskip 1em plus 0.5em minus
  0.4em\relax National Academies Press, 2020.

\bibitem{owidcoronavirus}
H.~Ritchie, E.~Mathieu, L.~Rodés-Guirao, C.~Appel, C.~Giattino,
  E.~Ortiz-Ospina, J.~Hasell, B.~Macdonald, D.~Beltekian, and M.~Roser,
  ``Coronavirus pandemic ({COVID}-19),'' \emph{Our World in Data}, 2020,
  \url{https://ourworldindata.org/coronavirus}.

\bibitem{owidhumandevelopmentindex}
M.~Roser, ``Human {D}evelopment {I}ndex ({HDI}),'' \emph{Our World in Data},
  2014, \url{https://ourworldindata.org/human-development-index}.

\bibitem{bishop2006pattern}
B.~Christopher, \emph{Pattern Recognition and Machine Learning}.\hskip 1em plus
  0.5em minus 0.4em\relax Springer, 2006.

\bibitem{jain1984quantitative}
R.~K. Jain, D.-M.~W. Chiu, and W.~R. Hawe, ``A quantitative measure of fairness
  and discrimination,'' \emph{Eastern Research Laboratory, Digital Equipment
  Corporation, Hudson, MA}, 1984.

\bibitem{lan2010axiomatic}
T.~Lan, D.~Kao, M.~Chiang, and A.~Sabharwal, ``An axiomatic theory of fairness
  in network resource allocation,'' in \emph{IEEE Conference on Information
  Communications (INFOCOM)}, 2010, pp. 1343--1351.

\bibitem{liu2021herd}
S.~Liu, Y.~Zhao, and Q.~Zhu, ``Herd behaviors in epidemics: A dynamics-coupled
  evolutionary games approach,'' \emph{arXiv preprint arXiv:2106.08998}, 2021.

\end{thebibliography}

\end{document}